\documentclass[showpacs,superscriptaddress,twocolumn,prb]{revtex4}
\usepackage{amsmath,graphicx}

\begin{document}

\title {Tuning the Josephson current in carbon nanotubes with the Kondo effect }
\author{A. Eichler}
\affiliation{Department of Physics, University of Basel,
Klingelbergstrasse 82, CH-4056 Basel, Switzerland}
\author{R. Deblock}
\affiliation{Laboratoire de Physique des Solides, Univ. Paris-Sud,
CNRS, UMR 8502, F-91405 Orsay Cedex, France}
\author{M. Weiss}
\affiliation{Department of Physics, University of Basel,
Klingelbergstrasse 82, CH-4056 Basel, Switzerland}
\author{C. Karrasch}
\affiliation{Institut f\"ur Theoretische Physik A and JARA --
Fundamentals of Future Information Technology, RWTH Aachen
University, 52056 Aachen, Germany}
\author{V. Meden}
\affiliation{Institut f\"ur Theoretische Physik A and JARA --
Fundamentals of Future Information Technology, RWTH Aachen
University, 52056 Aachen, Germany}
\author{C. Sch\"onenberger}
\affiliation{Department of Physics, University of Basel,
Klingelbergstrasse 82, CH-4056 Basel, Switzerland}
\author{H. Bouchiat}
\affiliation{Laboratoire de Physique des Solides, Univ. Paris-Sud,
CNRS, UMR 8502, F-91405 Orsay Cedex, France}


\begin{abstract}
We investigate the Josephson current in a single wall carbon
nanotube connected to superconducting electrodes. We focus on the
parameter regime in which transport is dominated by Kondo physics. A
sizeable supercurrent is observed for odd number of electrons on the
nanotube when the Kondo temperature $T_K$ is sufficiently large
compared to the superconducting gap. On the other hand when, in the
center of the Kondo ridge, $T_K$ is slightly smaller than the
superconducting gap, the supercurrent is found to be extremely
sensitive to the gate voltage  $V_{BG}$. Whereas it is
largely suppressed at the center of the ridge, it shows a sharp
increase at a finite value of $V_{BG}$. This increase can be
attributed to a doublet-singlet transition of the spin state of the
nanotube island leading to a $\pi$ shift in the current phase
relation. This transition is very sensitive to the asymmetry of the
contacts and is in good agreement with theoretical predictions.
\end{abstract}

\maketitle Metallic single wall carbon nanotubes have attracted a
lot of interest as 1D quantum wires combining a low carrier density
and a high mobility. Depending on the transparency of the interface
between the nanotube and the electrode, the conduction ranges from
insulating behavior and strong Coulomb blockade at low transparency
to nearly ballistic transport with conductance close to $4 e^2/h$
when the transparency is high \cite{liang01,babic}. The intermediate
conduction regime is particularly interesting, because in the case
of an odd number of electrons on the nanotube a strongly correlated
Kondo resonant state can form, where the magnetic moment of the
unpaired spin is screened by the spins of the electrons in the leads
\cite{gg-kondo}. Moreover, when the carbon nanotubes are in good
contact with superconducting electrodes, it is possible to induce
superconductivity and observe supercurrents, as was first
investigated in ungated suspended devices
\cite{kasumov99,kasumov03}. Proximity induced superconductivity was
then explored in gated devices with evidence of a strong modulation
of subgap conductance, but in most cases no supercurrent was
observed \cite{morpugo2000,buitelaar,lindelof06,eichler08}. More
recently, tunable supercurrents could be detected in the resonant
tunneling conduction regime, where the transmission of the contacts
approaches unity \cite{herero,sc_others}. In this regime, the
discrete spectrum of the nanotube is still preserved and the maximum
value of supercurrent is observed when the Fermi energy of the
electrodes is at resonance with ``the electron in a box
states" of the nanotube. A superconducting interference device was
fabricated with carbon nanotubes as weak links: supercurrent $\pi$
phase shifts occurred when the number of electrons in the nanotube
dot was changed from odd to even \cite{cleuziou07} corresponding to
the transition from a magnetic to a nonmagnetic state. Sharp
discontinuities in the critical current at this $0-\pi$ transition
in relation with the even-odd occupation number of the nanotube
quantum dot were also observed in single nanotube junction devices
\cite{lindelofnano07}. As pointed out in the SQUID experiment
\cite{cleuziou07} a Josephson current can be observed in the Kondo
regime, when the Kondo temperature $T_K$ is large compared to the
superconducting gap $\Delta$, confirming theoretical predictions
\cite{glazman,ra,belzig,yeyati,siano,karrasch08} and previous
experiments \cite{buitelaar,lindelofijp07} (with no determination of
supercurrents though).
\begin{figure}
\begin{center}
\includegraphics[clip=true,width=7.5cm]{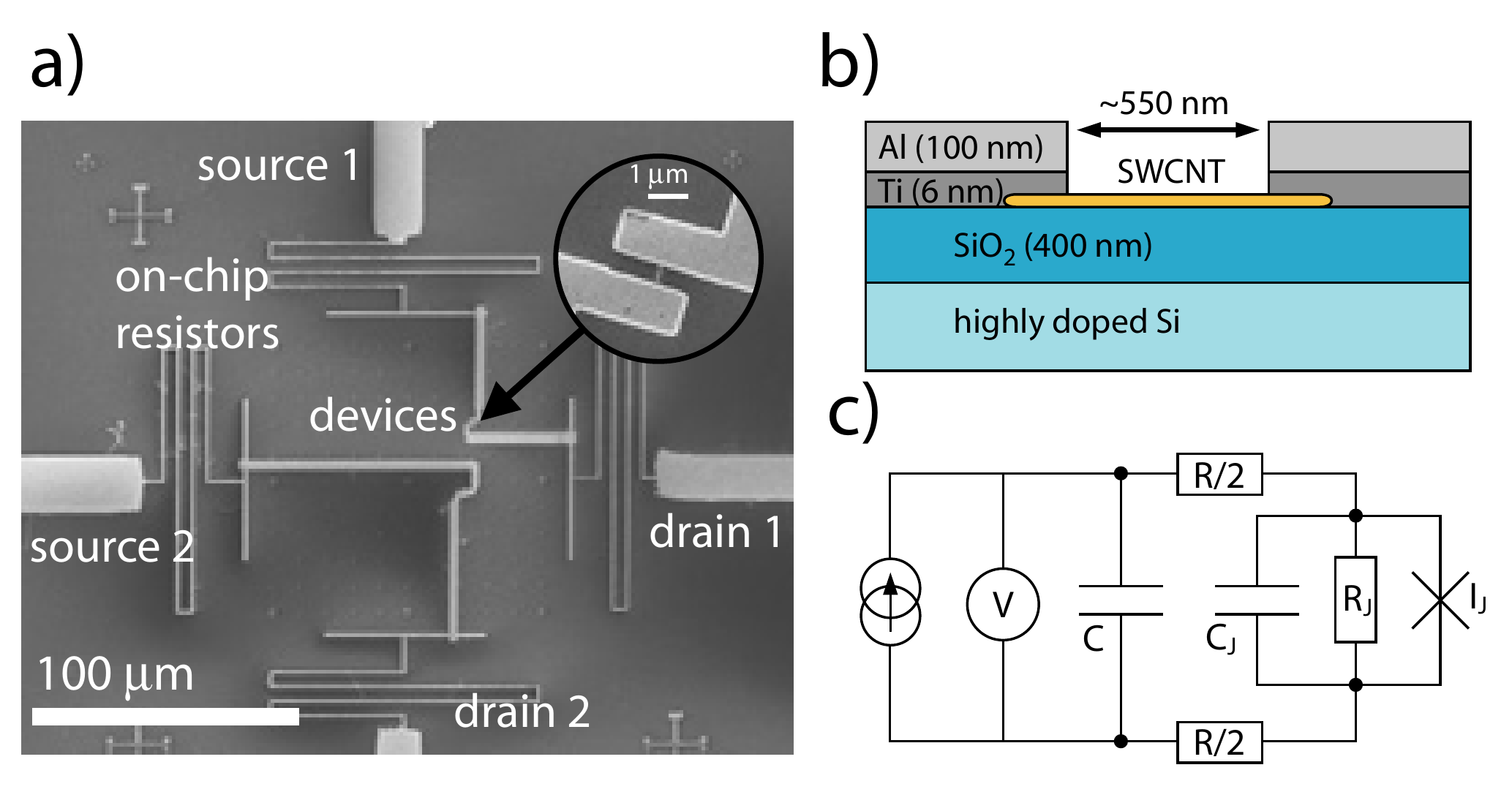}
\caption{a) Scanning electron micrograph of the electrical on-chip
environment of the nanotube. b) Cross section of the nanotube
with electrodes and backgate c) Schematic of the carbon nanotube
Josephson junction with its environment according to the extended
RCSJ model.
\label{fig1}}
\end{center}
\end{figure}

In the present work we explore in detail this competition between
Josephson and Kondo physics by monitoring on the same device as a function
of the gate voltage the bias dependence of the differential conductance in the normal state
and the Josephson current in the superconducting state. The value of this current is precisely determined by fitting the data with a theoretical model explicitly including the effect of the electromagnetic environment onto the junction. From the normal state data, we extract the
charging energy $U$ and the Kondo temperature $T_K$ as a function of
the gate voltage, and calculate the sum of the couplings to the
electrodes $\Gamma = \Gamma_L + \Gamma_R$ from the known form of
$T_K$ \cite{betheansatz}. The asymmetry $\Gamma_R / \Gamma_L$ is
determined from the value of the conductance on the Kondo resonance
for $T \ll T_K$. The value of this current is precisely determined by fitting the data with a theoretical model explicitly including the effect of the dissipative electromagnetic environment onto the junction. 
As expected a supercurrent is observed (for an odd
number of electrons on the nanotube) when  $T_K > \Delta$ and when
the asymmetry of the transmission of the electrodes is not too large
\cite{glazman,ra,belzig,yeyati,siano,karrasch08}.  On the other
hand, when  $T_K < \Delta$, the magnetic spin remains unscreened at
all temperatures leading to a $\pi$-junction with a very low
transmission of Cooper pairs \cite{oldworks}. We have  particularly
explored in this paper the intermediate regime where the Kondo
temperature can be tuned  with the gate voltage, on a  single Kondo
ridge, from a value slightly below  the superconducting gap at half
filling to a value $T_K > \Delta$. A sharp increase of the
supercurrent is  then observed, which is related to the transition
from a magnetic doublet state to a  non-magnetic singlet state of
the nanotube island. We compare these results with functional
renormalization group (FRG) calculations for the single impurity
Anderson model \cite{karrasch08} in a wide region of gate voltage. 

We grow single wall carbon nanotubes (SWCNT) by chemical vapor deposition on thermally
oxidized, highly doped silicon wafers. Individual SWCNTs are located
relative to predefined markers and contacted to Ti/Al leads using
electron-beam lithography (Fig. \ref{fig1}a,b). The Ti/Al leads are
superconducting below 1K. Measurements were done in a dilution
refrigerator with a base temperature of $T=40$mK. The cryostat was
equipped with a three stage filtering system consisting of LC
filters at room temperature, resistive micro-coax cables and finally
micro-ceramic capacitors in a shielded metallic box, which also
contained the samples and was tightly screwed onto the cold finger.
As additional filters against voltage fluctuations, resistors of the
order of 2k$\Omega$ were implemented on-chip as meanders in the Au
lines connecting the superconducting electrodes to the contact pads
(Fig. \ref{fig1}a). Measurements were done with a lock-in technique
with either a AC-voltage of 5  to 10 $\mu$V, measuring differential conductance
$dI/dV$, or  a  AC-current of $10pA$, measuring differential
resistance $dV/dI$, both as a function of an additional DC-bias
voltage or current, respectively. Fig. \ref{fig1}c shows a schematic
view of our device. It consists of a Josephson junction with current
$I_{J} (\phi)$ parallel to a capacitor $C_{J}$ and a shunt
resistance $R_{J}$. We obtain a rough estimate of $C_{J}\approx
100$aF from the charging energy $U=e^2/2C_{J}$. The junction
resistor $R_J$  represents the contribution of quasi-particles to
dissipation at frequencies of the order of the plasma frequency of
the junction $\omega_P$. The outer capacitance $C$ represents the
capacitances of the leads to ground within the electromagnetic
horizon of the junction, which we estimate as $2\pi c / \omega_p$ to
some centimeters, depending on $R_J$. It is therefore mainly
determined by the capacitance of the metallic leads to the highly
doped backgate, and can be estimated to $C \approx 8$pF
\cite{dissipationbackgate}. The series resistances $R$ represent the
lithographically defined on-chip resistors. This situation
corresponds to the ``extended resistively and capacitively shunted
junction'' (extended RCSJ) model \cite{RCSJ, herero,
lindelofnano07,ambegaokar}, which we will use later.
\begin{figure}
\begin{center}
\includegraphics[clip=true,width=7.5cm]{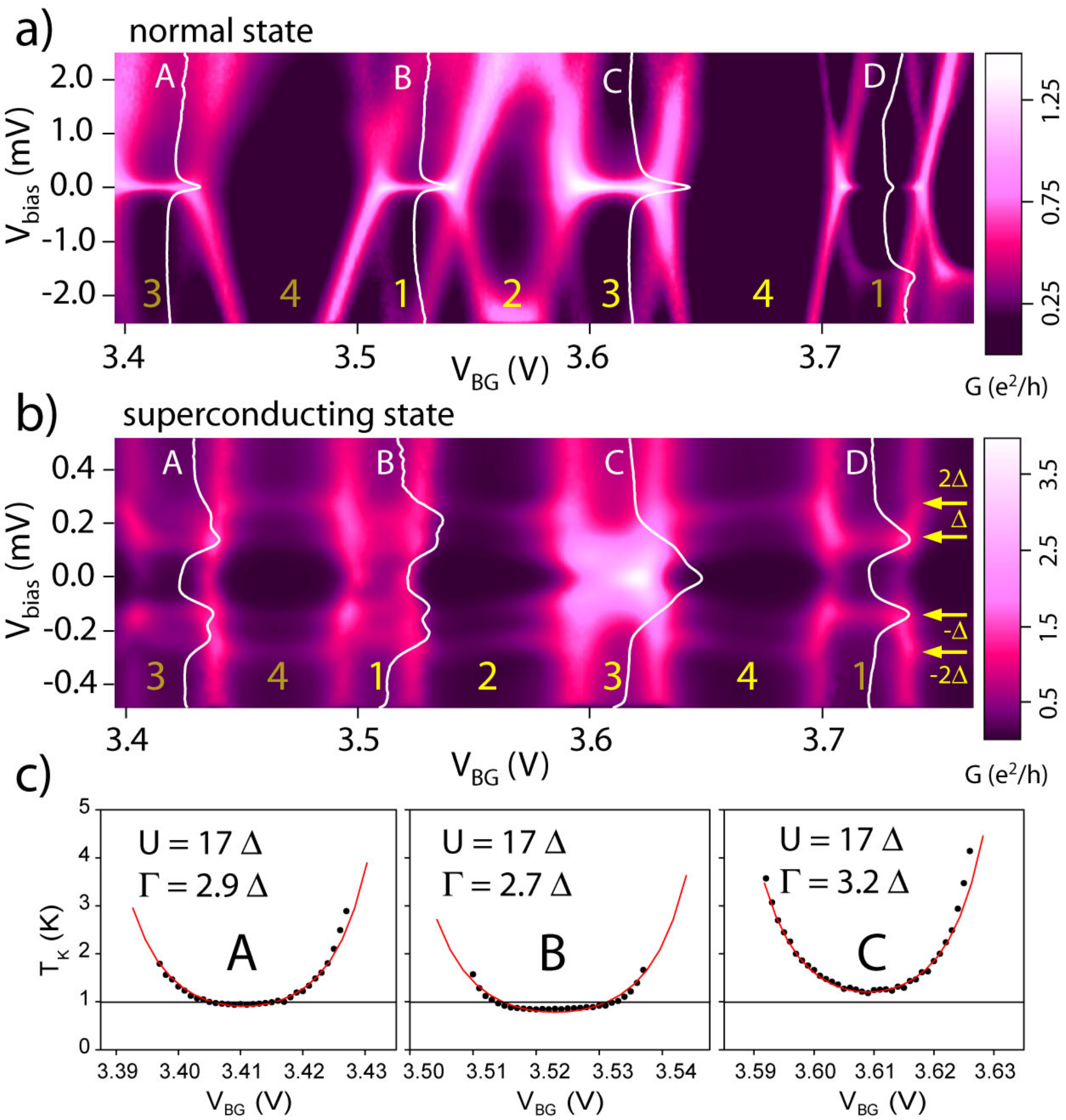}
\caption{Color scale plot showing gate and voltage bias dependence
of the differential conductance. a) normal state. b) superconducting
state. White lines correspond to vertical sections of $G$ versus
$V_{bias}$ in the middle of the four Kondo ridges labeled A to D. c)
gate dependence of the Kondo temperature extracted on the Kondo
ridges far from the degeneracy points. Red lines are fits to eq.
\ref{TK}, the black line corresponds to $T_C = 1.0K$. $U$ and
$\Gamma$ are given in units of $\Delta$. \label{fig2}}
\end{center}
\end{figure}

We first characterize the CNT-quantum dot with the electrodes
driven normal by a small magnetic field ($100$mT). The color scale
plot of $dI/dV$ as a function of $V_{bias}$ and $V_{BG}$ (the
``charge stability diagram"),   displays a regular sequence of
``Coulomb blockade diamonds" over a wide range of $V_{BG}$. In
other regions this pattern is replaced by a smoother gate
dependence characteristic of Fabry-Perot oscillations
\cite{liang01}. For further analysis, we concentrate on the gate
voltage region between 3 and 4 Volts  (see Fig. \ref{fig2}a). It
shows a fourfold periodicity in the size of the Coulomb blockade
diamonds \cite{fourfold}, which indicates a clean nanotube with
the twofold orbital degeneracy of the electronic states preserved.
We extract $U = 2.5 \pm 0.3$meV as estimated from the size of
Coulomb blockade diamonds with an odd number of electrons. In all
states with odd occupation, the Kondo effect manifests itself
through a high conductance region around zero bias, the so called
Kondo ridge. The Kondo temperature $T_{K}$ can be estimated from
the half width of the peaks of these lines which can be fitted by
Lorentzian curves \cite{GG}. The Kondo temperature goes through a
minimum of the order of 1K in the center of the ridges and
increases on the edges. It is possible to follow this gate
dependence along the Kondo ridges A to C (Fig. \ref{fig2}c). The
intensity on ridge D is too weak for such an analysis. $T_K$ can
be well fitted by the expression predicted by the Bethe Ansatz
\cite{betheansatz}:
\begin{equation}
T_K=\sqrt{U\Gamma/2}\exp\left[- \frac{\pi}{8U
\Gamma}|4\epsilon^2-U ^2|\right] \label{TK}
\end{equation}
where $\epsilon$ is the energy shift measured from the center of the
Kondo ridge. Taking $U$=2.5meV as determined above, the value of
$T_K$ at $\epsilon =0$ leads to the characteristic coupling energies
$\Gamma=\Gamma_{R}+ \Gamma_{L}$ between the electrodes for each
Kondo ridge (Fig. \ref{fig2}c). The gate voltage $V_{BG}$ dependence
of $T_K$ yields the ratio $\alpha$ between the electrostatic energy
$eV_{BG}$ and the Fermi energy of the nanotube $\epsilon$ equal to
$20\pm 1$. This value agrees to within 20\% with the value deduced
from the normal state conductance data. This fit based on the single
level Anderson impurity model is only valid when $U/\Gamma$ is
sufficiently large and $|\epsilon| \ll U$. Since for all peaks $T
\ll T_{K}$, the maximum conductance of the ridges yields the
asymmetry of the coupling $\Gamma_{R}/\Gamma_{L}$,  6.8, 6.2, 2.5
and 70 for A, B, C and D, respectively.
\begin{figure}
\begin{center}
\includegraphics[clip=true,width=7.5cm]{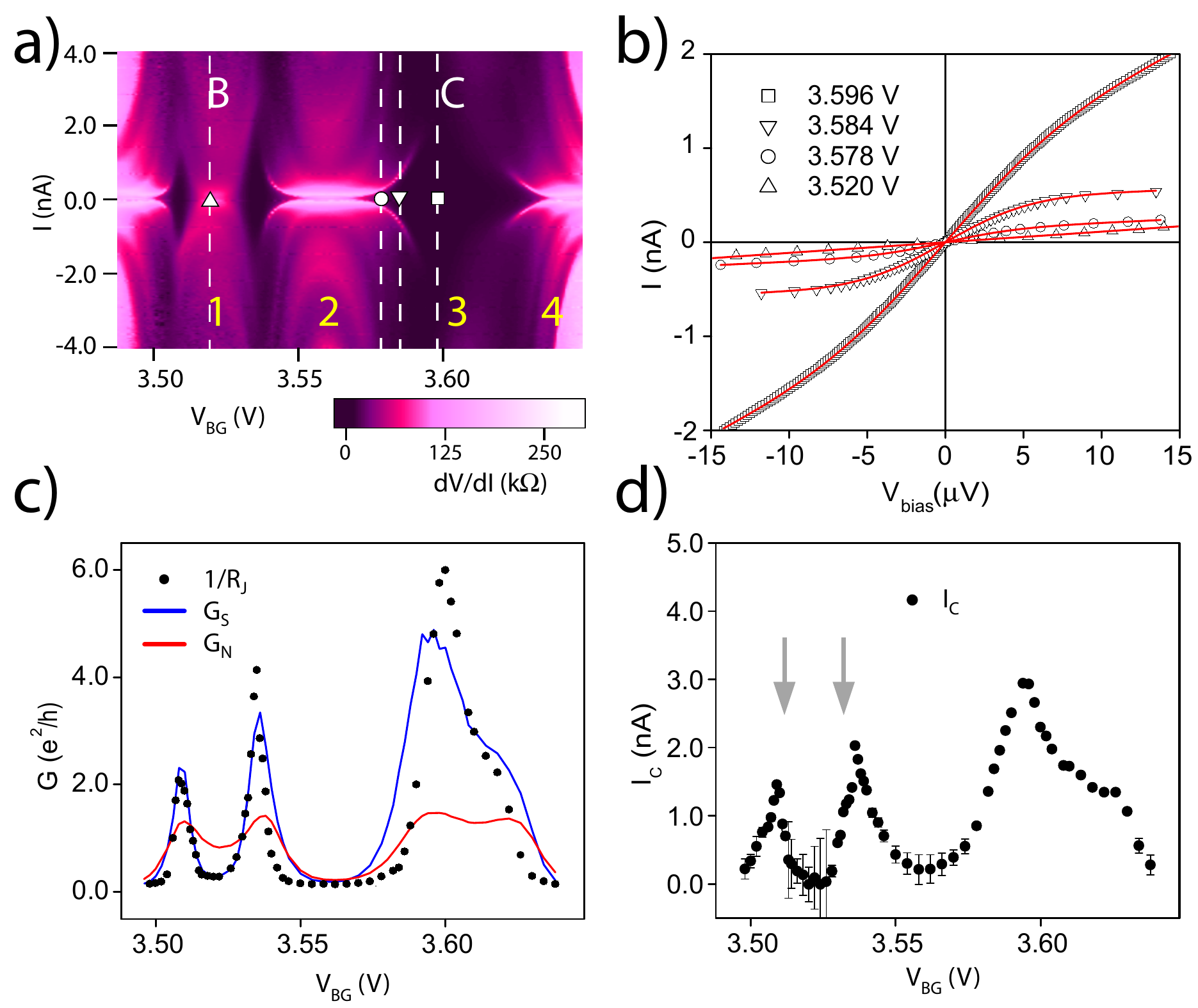}
\caption{a) Color scale plot of the differential resistance as a
function of DC bias current and backgate voltage. b) I-V
characteristics for different backgate voltages. The red lines are
fits using equation \ref{IV}. c) Conductances $G_J=1/R_J$ (black
circles), $G_S$ (blue line) and $G_N$ (red line). d) Gate voltage
dependence of the critical current extracted from fits to eq.
\ref{IV}. \label{fig3}}
\end{center}
\end{figure}

By switching off the magnetic field, we allow the leads
 to become superconducting. Fig. \ref{fig2}b shows $dI/V$ for the same gate voltage range as in Fig.
\ref{fig2}a but for a smaller bias voltage range. The BCS-like
density of states in the electrodes leads to new features in the
stability diagram, which are horizontal lines at $V_{bias} = \pm 2
\Delta_{0}$ and $\pm \Delta_{0}$ due to the onset of
quasiparticle tunneling and Andreev reflection, respectively.  We
can derive the value of the superconducting gap $\Delta_{0} =0.15
\pm 0.02$meV, which corresponds well to the expected $T_C \simeq 1K$
for the bilayer Ti/Al. Although they have similar Kondo-temperatures
in the normal state, the Kondo-ridges A, B, C and D show a very
different behavior when the electrodes are superconducting.
 The Kondo ridges A, B and D are suppressed or reduced in
amplitude by superconductivity, reflecting that $T_K < T_C$  in
the center of the Kondo ridge. Contrary to ridge A,B and D, that
show a minimum at zero bias,  there is a strong enhancement of
conductance in ridge C, with $G$ reaching a value roughly four times
larger than in the normal state. Note that the states A, C and D
show an enhancement of conductance at $V_{bias} = \Delta/e$, to
values larger than the conductance at $V_{bias} = 2\Delta/e$. This
enhancement of the first Andreev process is due to the "even-odd"
effect in Andreev transport, that has recently been described in
\cite{sand-jespersen08,eichler08}.

For a measurement of the supercurrent the device has to be current
biased. We simultaneously use AC and DC bias while measuring the
resulting voltage drop across it. From the AC part, we obtain data
on the differential resistance (Fig. \ref{fig3}a). By numerical
integration we get I-V curves that show a supercurrent branch and a
smooth transition to a resistive branch with higher resistance (Fig. \ref{fig3}b). The transition between the two regimes is not
hysteretic, and the supercurrent part exhibits a non zero resistance
$R_S$ at low bias even if we subtract the contribution of the
on-chip resistances $R$. This behavior is common in mesoscopic
Josephson junctions that have a high normal state resistance of the
order of the resistance quantum $h/e^2$. To extract the supercurrent, we
use a theory that explicitly includes the effect of the
dissipative electromagnetic environment onto the junction in the frame of the
already mentioned extended RCSJ-model \cite{herero,lindelofnano07,ambegaokar}.
Using the external resistor $R$ \cite{overdamped} (Fig. \ref{fig1}c)
and temperature $T$ as input parameters, we can thus extract the critical current $I_{c}$
and the junction resistance $R_{J}$, for every measured backgate
voltage, from a fit to~:
\begin{equation}
I(V_{bias}) = \left\{ I_c  Im \left[
\frac{I_{1-i\eta}(I_c \hbar/2 e k_B
T)}{I_{-i\eta}(I_c \hbar/2 e k_B T)} \right]
 + \frac{V_{bias}}{R_j} \right\} \frac{R_j}{R_j+R} \label{IV}
\end{equation}
where $\eta=\hbar V_{bias}/2 e R k_B T$ and $I_{\alpha}(x)$ is the
modified Bessel function of complex order $\alpha$
\cite{lindelofnano07}. The resulting values are plotted as a
function of $V_{BG}$ in Figs. \ref{fig3}c and d. The junction
conductance $G_J = 1/R_J$ relates well to the differential
conductance $G_S = 1/R_{S}$ extracted from the AC part of the
current biased data, especially around the resonance degeneracy
points where the conductance is high. The normal state conductance
$G_{N}$ deviates from $G_{S}$ most notably in the states B and C,
where Kondo physics plays a key role. In state B, $G_{N} > G_{S}$
whereas the ridge in state C persists in the superconducting state,
resulting in a further enhancement of the conductance.
\begin{figure}
\begin{center}
\includegraphics[clip=true,width=7.5cm]{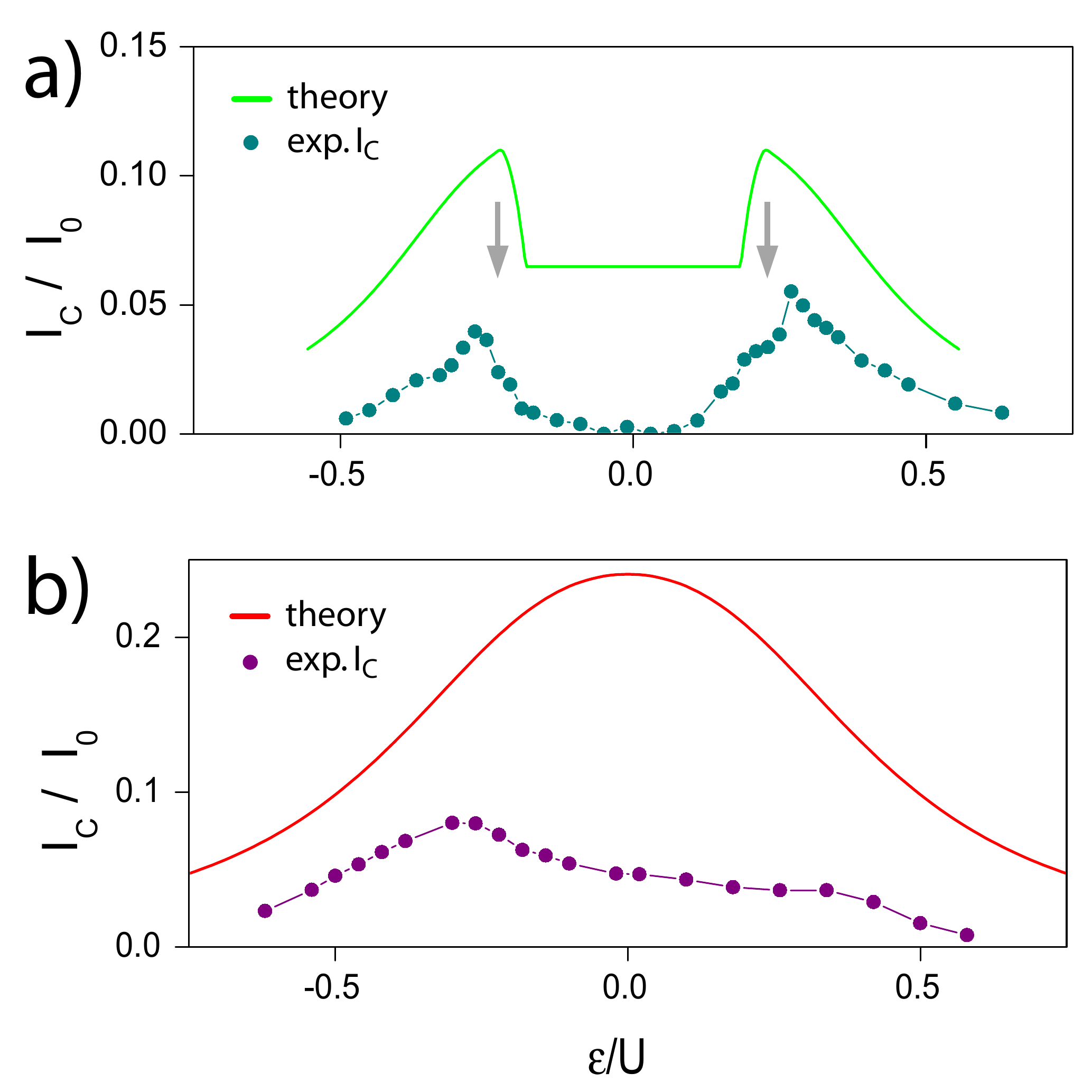}
\caption{Comparison with FRG calculations: Calculated
$I_c(\epsilon)$ in units of $I_0=e\Delta/\hbar$ for $\Gamma=2
\Delta, \Gamma/U=0.11, \Gamma_R/\Gamma_L=6$ in a) and $\Gamma=2
\Delta, \Gamma/U=0.2, \Gamma_R/\Gamma_L=3$ in b). Experimental data
for $I_c(\epsilon)$ are shown for Kondo ridge B in a) and Kondo
ridge C in b). \label{fig4}}
\end{center}
\end{figure}

The supercurrent exhibits peaks at the maximum values of the
conductance (Fig. \ref{fig3}d). Its behavior between peaks varies
strongly in states with even and odd occupation. In the following
we focus on the Kondo ridges B and C. Whereas the supercurrent on
ridge C varies nearly proportionally to the conductance, we
observe a sharp drop of supercurrent on ridge B as illustrated on
Fig. \ref{fig3}d (arrows). This can be understood considering that
$T_K$ is smaller on ridge B than on ridge C (see Fig. \ref{fig2}c). As a result the magnetic moment of the excess
electron on the nanotube remains unscreened in the superconducting
state near $\epsilon=0$ which is not the case for ridge C.
At this stage we compare our experimental findings to approximate
zero-temperature FRG calculations which allow for extracting both
the stability regions of the  screened singlet and magnetic
unscreened doublet phases and the complete supercurrent-phase
relation $I_J(\phi)$ within a model of an Anderson impurity coupled
to two superconducting electrodes \cite{karrasch08}. It was
previously observed that the Josephson current cannot be simply
described as a single function of the ratio $T_K/\Delta$
\cite{glazman}. The relevant parameters are the on site Coulomb
repulsion energy $U$, the level position $\epsilon$ related to $V_{BG}$, and
 the transmission of the electrodes $\Gamma_R$ and $\Gamma_L$
compared to the superconducting gap. In Fig. \ref{fig4} we present
a comparison between our experimental data to the theoretical
predictions for $I_c$ (defined as the maximum value of
$|I_J(\phi)|$ over $\phi$), as a function of the position in
energy of the Anderson impurity level, for similar values of
theoretical and experimental parameters. It is remarkable that the
gate dependence of the critical current on the Kondo ridges can be
qualitatively described. That is in particular, the existence of a
singlet, doublet ($0/\pi$) transition on ridge B at nearly the
right value of gate voltage when renormalized with the charging
energy $U$ (arrows), and the absence of a ($0/\pi$) transition for
ridge C corresponding to higher values of $T_K$. One point of
disagreement between theory and experiment concerns the amplitude
of the critical current in the doublet state ($\pi$ junction)
region (around $V_{BG}= 3.65$V). $I_c$ is theoretically found to
be reduced by only a factor of two compared to its value in the
singlet region whereas basically no trace of superconductivity
could be detected experimentally. It is known however, that the
$\pi$-phase current computed from the approximate FRG is too large
compared to numerical renormalization group (NRG) data, which are known 
to be more accurate but only available at the
center of the Kondo ridge \cite{karrasch08}.

Finally, let us emphasize the importance of the asymmetry of the
transmission of the electrodes which tends to reduce considerably
the supercurrent. This is particularly striking for the data on
ridge D, for which $\Gamma_R / \Gamma_L =70$, where no supercurrent
could be measured in spite a value of $T_K$ of the order of 1K. This
reduction of $I_c$ by the asymmetry of contacts is also found in FRG
calculations \cite{karrasch08}. It is moreover accompanied by  a
modification of the stability regions for the singlet (screened) and
doublet (magnetic) states which strongly depend on the phase
difference between the superconducting electrodes. The
non-magnetic singlet state is stabilized with respect to the
magnetic doublet state in the vicinity of $\phi=\pi$, which results
in a strong modification of the current phase relation which
unfortunately cannot be checked in a single $I_c$ measurement
\cite{karrasch08}.

In conclusion, we have shown that it is possible to tune the
amplitude of the supercurrent in a carbon nanotube Josephson
junction within a Kondo ridge in a very narrow range of gate
voltage. Our data are in good qualitative agreement with
theoretical findings and should stimulate new experiments where the
whole current phase relation is measured \cite{huard}. We
acknowledge fruitful discussions with Y. Avishai, S. Gu\'eron, W.
Belzig, A. Levy-Yeyati, T. Novotn\'y, J. Paaske, and B.
 R\"othlisberger. This work was supported by the
EU-STREP program HYSWITCH and by the Deutsche Forschungsgemeinschaft
via FOR 723 (CK, VM).


\begin{thebibliography}{99}
\bibitem{dissipationbackgate}
Due to this large value of $C$ it is possible  that  the finite
resistivity of the backgate also contributes to dissipation at very
high frequency.
\bibitem{overdamped} Note the absence of explicit dependence on the
capacitances $C$ and $C_j$ for an overdamped junction, as it is the
case in our experiment (Q=0.03).
\end{thebibliography}
\end{document}